\documentclass[a4paper,12pt,onecolumn,superscriptaddress]{revtex4}

\usepackage{graphicx}
\usepackage{amsmath}
\usepackage{latexsym}
\usepackage{epstopdf}

\newcommand{\beq}{\begin{equation}}
\newcommand{\eeq}{\end{equation}}

\begin{document}

\title{GW approach to Anderson model out of equilibrium: Coulomb blockade and 
false hysteresis in the I-V characteristics}

\author{Catalin D. Spataru}
\affiliation{Center for Electron Transport in Molecular Nanostructures and 
Center for Integrated Science and Engineering, 
Columbia University, New York, NY 10027, USA}
\affiliation{Sandia National Laboratories, Livermore, CA 94551, USA}
\author{Mark S. Hybertsen}
\affiliation{Center for Functional Nanomaterials, Brookhaven 
National Laboratory, Upton, NY 11973, USA}
\author{Steven G. Louie}
\affiliation{Department of Physics, University of California at Berkeley,
Berkeley, CA 94720, USA}
\affiliation{Materials Sciences Division, Lawrence Berkeley National
Laboratory, Berkeley, CA 94720, USA}
\author{Andrew J. Millis}
\affiliation{Department of Physics, Columbia University, New York, 
NY 10027, USA}

\begin{abstract}
The Anderson model for a single impurity coupled to two
leads is studied using the $GW$ approximation in the strong 
electron-electron interaction regime 
as a function of the alignment of the 
impurity level relative to the chemical potentials in the
leads. We employ a non-equilibrium Green's function technique to calculate
 the electron self-energy, the spin density and
 the current as a function of bias across the junction. 
In addition we develop an expression for the change in the expectation value of the energy
of the system that results when the impurity is coupled to the leads, 
including the role of Coulomb interactions through the electron
self energy in the region of the junction.
The current-voltage characteristics 
calculated within the GW approximation exhibit
Coulomb blockade. Depending on the gate voltage 
and applied bias, we find that there can be more than one 
steady-state solution 
for the system, which may give
rise to a hysteresis in the I-V characteristics.
We show that the hysteresis is an artifact of the $GW$
approximation and would not survive if quantum fluctuations beyond
the $GW$ approximation are included.
\end{abstract}

\maketitle

\newpage
\section{Introduction}
Transport through nanoscale junctions poses a number of interesting physical
problems.
In particular, electron-electron 
interaction effects may be important, as evidenced by the observation of
phenomena such as the Coulomb blockade and the Kondo 
effect \cite{McEuen,Park}.
The local electronic structure is also important.
The energy and character of the electronic states in the junction region 
that are responsible for 
electron transport will depend on the details of bonding between
the molecule and the electrode.
This has motivated the use of
ab initio theories for electron transport through
nanostructures that are based on Density Functional Theory (DFT).
However, local
density functionals do not treat the discreteness of charge
properly \cite{Natelson,Burke}. In particular Coulomb blockade
phenomena become problematic. 
Even on the level of model systems, a complete solution of the
nonequilibrium interacting electron problem is not available. The
numerical methods which work so well in equilibrium 
are only beginning to be applied to non-equilibrium systems
\cite{Anders08,Weiss08,Al-Hassanieh06,Kirino08,Muehlbacher08,Schiro08,Schmidt08,Werner08}.
Many groups are
exploring selfconsistent perturbative and other, nonperturbative approaches
\cite{Ferretti,Darancet,Thygesen,Thygesen07b,Thygesen08,Mitra05,Mitra07,Segal07}
However, a complete treatment which can be extended to incorporate actual,
junction-specific aspects is not yet available. 

In this work, we study a model system, namely the single impurity 
Anderson model \cite{Anderson} coupled to two leads.
We use a Green's function approach to calculate the properties of the junction,
both in equilibrium and as a function of applied bias across the junction.
The electron-electron interactions are incorporated through the electron self energy
operator on the impurity, using an out-of-equilibrium generalization 
of the $GW$ approximation \cite{Hedin}.
Using this approach we can calculate the local spin density in the junction
and the current as a function of bias.
In addition we develop and apply an extension to non-zero bias of the
usual expression \cite{KSS} for the change in the average
energy of the impurity due to coupling to the leads.
The $GW$ approximation has been widely and successfully used to study electronic
excitations in materials at equilibrium with a realistic, atomic scale description
\cite{Hybertsen,Aryasetiawan,Aulbur00,Stan06,Schilfgaarde06}.
This is one of the motivations to study the 
out-of-equilibrium generalization for nanoscale junctions \cite{Darancet,Thygesen,Thygesen07b,Thygesen08}.
In particular, the intermediate coupling/interaction regime of the single impurity Anderson
model has recently been studied using the $GW$ approximation \cite{Thygesen,Thygesen07b}.

We are interested in the intermediate to strong coupling regime,
in which Coulomb blockade effects are important.
At equilibrium and for zero temperature,
as the local Coulomb interaction on the impurity is increased (relative
to the hybridization with the leads) a local moment forms.   In the
limit of $k_BT\rightarrow 0$ and vanishing bias,
the local moment on the impurity is quenched through formation of a singlet
ground state. The spectral function splits into three parts, two
Hubbard bands and
one central Kondo peak.  
In a closely related earlier study \cite{Wang}, it was shown that
in the regime of intermediate strength of the Coulomb interaction,
the $GW$ approximation 
provides an incorrect representation of the linear response
conductance. In fact, this regime is not well
described at equilibrium even by more sophisticated perturbative 
approaches, such as the fluctuation-exchange approximation \cite{White,Flex}. 
Here we probe the strong coupling, Coulomb blockade regime. In this regime, 
the Kondo temperature $T_K$ becomes very small and at
experimentally relevant temperature scales the Kondo peak will be washed out.
Similarly, when considering 
bias 
large
compared to the Kondo temperature, the Kondo peak also gets washed out
\cite{MeirWingreenLee,Hershfield}. In these regimes
a self-consistent perturbative approach may be adequate.
We find through non-equilibrium calculations 
that the self-consistent $GW$ approximation
can describe important features of the Coulomb blockade regime, such
as the Coulomb diamond signature with no Kondo-assisted tunneling, 
in accordance with experiments on single-molecule transistors
characterized by weak effective coupling between molecule and electrodes
\cite{McEuen}.

The non-equilibrium
GW
calculations exhibit hysteresis in the IV characteristics:
at some values of applied bias and gate voltage, there is more than
one steady state solution. A related example of bistability has been
found in DFT calculations of a junction involving an organometallic
molecule \cite{Baranger}. However, we believe that in the problem that
we study here, the hysteresis is an artifact of the approximation 
\cite{Mitra05}. In fundamental terms, a molecular junction is a
quantum field theory in $0$ space and $1$ time dimension. Model 
system calculations \cite{Mitra05,Mitra07} have confirmed that 
departures from equilibrium act as an effective temperature which 
allows the system to explore all of its phase space,
preventing bistability from occurring.  We will show by an energy 
calculation that in the present problem similar processes exist.

The rest of the paper is organized as follows. In Section II, the model 
Hamiltonian is described.  
Section III presents the non-equilibrium, self-consistent Green's 
function approach that we use,
including the GW approximation, an expression for the change in the
average energy as well as an expression for the current that allows to
distinguish the Landauer-like and the non-coherent contributions.
The results of the calculations for
the single impurity Anderson model are developed in Section IV.
Derivations of the expressions for the physical observables appear in 
Appendices A, B and C.

\section{Model Hamiltonian}
We consider the Anderson model for an impurity coupled
symmetrically to non-interacting leads. 
We are interested in 
steady-state solutions of this system.
The Hamiltonian
describing the system, $H$, can be written as a sum of a non-interacting
part, $H_0$, plus an interacting one, $H_{e-e}$, describing the
electron-electron interaction in the impurity: $H=H_0+H_{e-e}$.

\begin{figure}
\resizebox{12.0cm}{!}{\includegraphics{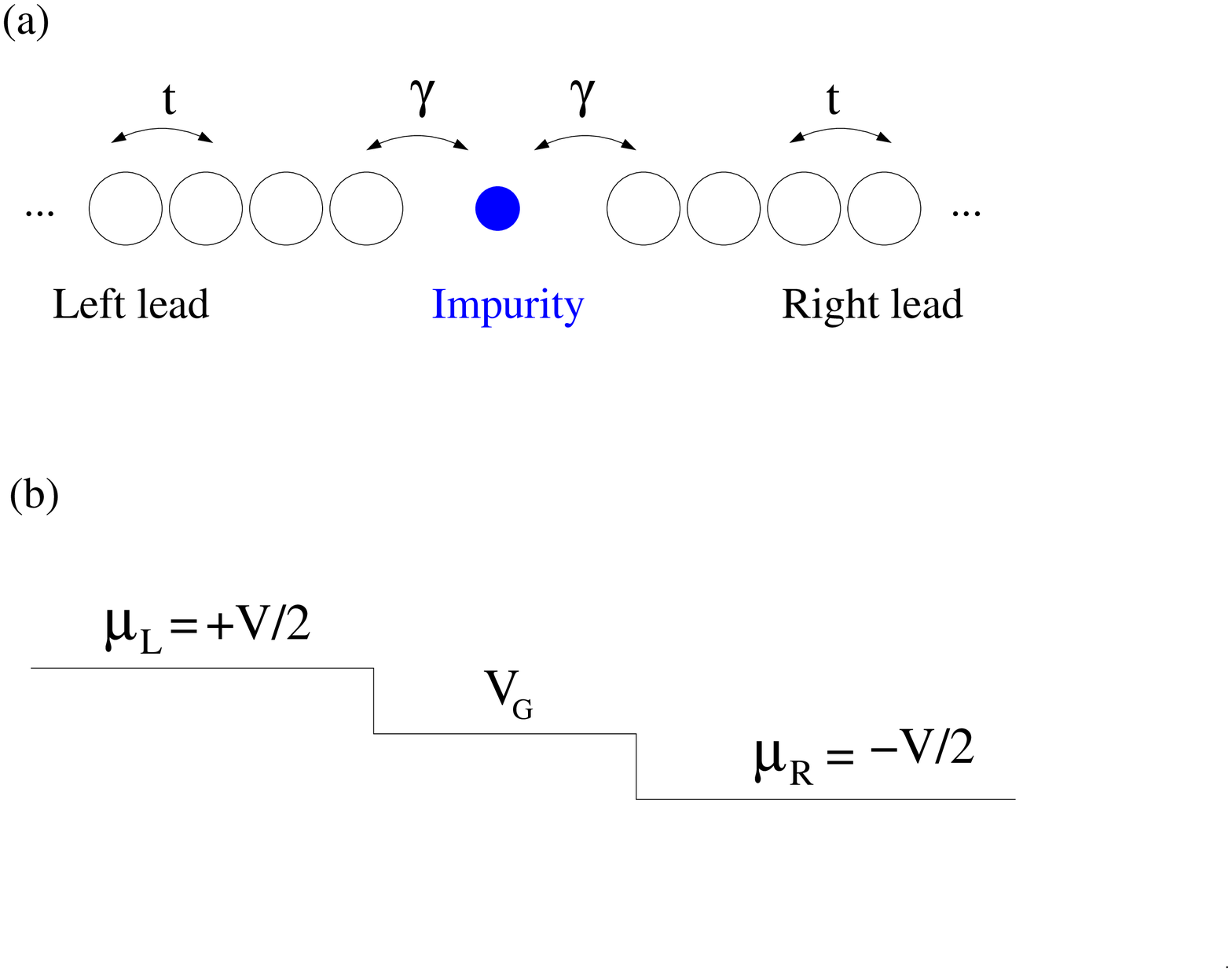}}
\caption{Schematic view of the Anderson impurity model system considered.
(a) Tight binding model for the non-interacting system.
(b) Definition of applied source-drain bias $V$ and gate voltage $V_G$. }
\label{model}
\end{figure}

The non-interacting part is treated at the tight-binding level (Fig. \ref{model}a). 
The left (L) and right (R) leads are modeled as semi-infinite chains
of atoms (i=1,... $\infty$ or -1,... $-\infty$), characterized by the 
hopping parameter t and chemical
potentials $\mu_L$ and $\mu_R$. 
We choose $t=5$, resulting in the
band-width of the metallic leads extending to $\pm 10$ about the 
chemical potential of each lead
which we fix at the center of each electrode band.
The system is driven out of
equilibrium by applying a source-drain bias voltage V, setting $\mu_L=-\mu_R=V/2$;
the impurity levels can also be shifted according to a gate voltage $V_G$ (Fig. \ref{model}b). 
The hybridization term describes the coupling between the  
impurity (site $0$) and the nearest atoms of the two leads (sites
$\pm 1$), and is parameterized according to the hoping parameter
$\gamma$. 
\beq
H_0=\mu_L~N_L+
\mu_R~N_R +V_G~n_0
-t (\sum_{i=-\infty}^{-2}+\sum_{i=1}^{\infty}) \sum_{\sigma} 
   (c_{i\sigma}^\dagger c_{i+1\sigma}
     +c_{i+1 \sigma}^\dagger c_{i \sigma})
-\gamma \sum_{i=-1,1} \sum_{\sigma} 
   (c_{i\sigma}^\dagger c_{0 \sigma}
     +c_{0 \sigma}^\dagger c_{i \sigma})
\eeq
where $N_{L(R)}$ are the electron number operators in the $L(R)$ leads:
\beq
N_L=\sum_{i=-\infty}^{-1}\sum_\sigma
c_{i\sigma}^{\dagger}c_{i\sigma};~~~~
N_R=\sum_{i=1}^{\infty}\sum_\sigma
  c_{i\sigma}^{\dagger}c_{i\sigma}
\eeq
and $n_0$ is the electron number in the impurity:
\beq
n_0=\sum_{\sigma} c_{0\sigma}^{\dagger}c_{0\sigma}
\eeq

The electron-electron interaction inside the 
impurity is taken into account through the usual U-term:
\beq
H_{e-e}=U~n_{0 \uparrow} n_{0 \downarrow} =\frac{1}{2}
\sum_{\alpha,\alpha',\beta,\beta'} c^{\dagger}_{0,\alpha}
c^{\dagger}_{0,\beta} \tilde{V}_{\alpha  \alpha',\beta \beta'}
c_{0,\beta'} c_{0,\alpha'}
\label{Hamilt}
\eeq
There are several choices we can make for the 2-particle interaction 
$\tilde{V}_{\alpha  \alpha',\beta \beta'}$. We choose one that
describes non-spin-flip scattering:
\beq
\tilde{V}_{\alpha  \alpha',\beta \beta'}=V_{\alpha \beta}~\delta_{\alpha \alpha'}~\delta_{\beta \beta'}
\eeq
and has a spin-dependent form:
\beq
V_{\alpha \beta}=U~(1-\delta_{\alpha \beta})
\eeq
Another choice for the 2-particle interaction, which results in the
same Hamiltonian as in Eq. \eqref{Hamilt}, would be one with a
spin-independent form: $V_{\alpha \beta}=U$. However,
in the context of the $GW$ approximation for the Anderson model, 
the spin-dependent form is a better choice \cite{Wang}. Indeed,
it has been shown that the
spurious self-interactions can be a major source of error 
in transport calculations, especially when the coupling to the leads is 
weak \cite{Burke}. Comparing the two choices for $V_{\alpha \beta}$,
the spin-dependent one has the advantage of being 
free of self-interaction effects, and it also accounts for more
quantum fluctuations in the spin-spin channel \cite{Wang}.

In the present model, 
the potential due to the applied source-drain bias $V$ and gate voltage $V_G$ 
changes only at the junction contacts (Fig. \ref{model}b). 
Also, the direct electron-electron 
interaction between the impurity and the leads 
is neglected. 
These approximations are justified in realistic systems in which the
screening length in the leads is very short.
We shall be interested in the limit of very small effective coupling
to the leads
$\Gamma\equiv2\gamma^2/t$. Our choices $\gamma=0.35$ and $t=5$ imply 
$\Gamma=0.05$. The on-site Coulomb repulsion between a
spin-up and a spin-down impurity electron is set to $U=4.78 \simeq 100
\Gamma$.
At equilibrium and
half-filling, the Kondo temperature $T_K$ is then \cite{Haldane}:
\beq
T_K \approx 0.2\sqrt{2\Gamma U}\exp(-\pi U/8\Gamma)
\eeq
which is thus negligible small.
The results that we
present for this set of parameters hold, qualitatively, for a wide
range of parameters consistent with a weak hybridization and strong
Coulomb interaction regime.

\section{Self-consistent non-equilibrium Green's function formalism}
\label{section_formalism}
\subsection{Hamiltonian and Basic Formalism}
Electron correlation effects in the impurity are studied 
using a non-equilibrium Green's function
formalism, by solving self-consistently for the various [retarded (r),
advanced (a), lesser ($<$) and greater ($>$)] Green's
functions of the impurity \cite{Haug,Datta}:
\beq
G^r(\omega)=[(\omega-V_G) I -\Delta_L^r(\omega)-\Delta_R^r(\omega)-V^H-\Sigma^r(\omega)]^{-1}
\eeq
\beq
G^<(\omega) = G^r(\omega)[if_L(\omega)\Gamma_L(\omega)+if_R(\omega)\Gamma_R(\omega)+\Sigma^<(\omega)]G^a(\omega)
\label{Gless}
\eeq
where all quantities are matrices in the space spanned by the
junction degrees of freedom, in the present case the up and down components of
the impurity spin \cite{notation}.

Above, $\Delta^r$ stands for the retarded lead self-energy, which, for
our model Hamiltonian, takes the form \cite{Budau}:
\begin{eqnarray} 
\Delta^r_{L(R)}(\omega)
         =I~\frac{\gamma^2}{2t^2}
     \left [\omega-\mu_{L(R)}-\sqrt{(\omega-\mu_{L(R)})^2-4t^2}\right ] ,& \omega-\mu_{L(R)} >2t
	 \nonumber\\
	 =I~\frac{\gamma^2}{2t^2}
     \left [\omega-\mu_{L(R)}-i\sqrt{4t^2-(\omega-\mu_{L(R)})^2}\right ] , & |\omega-\mu_{L(R)}| \leq 2t
	 \nonumber\\
	 =I~\frac{\gamma^2}{2t^2}
     \left [\omega-\mu_{L(R)}+\sqrt{(\omega-\mu_{L(R)})^2-4t^2}\right ] ,  &\omega-\mu_{L(R)} < -2t
\label{lead_sigma}
\end{eqnarray}
and we have used the notation:
\beq
\Gamma_{L(R)}(\omega)\equiv
i[\Delta_{L(R)}^r(\omega)-\Delta_{L(R)}^r(\omega)^{\dagger}].
\label{Gama}
\eeq
The hybridization functions $\Delta_{L(R)}$ are centered on the
chemical potentials $\mu_{L(R)}$, such that the isolated leads are neutral.

$V^H$ represents the Hartree potential:
\beq
V^H_{\sigma\sigma'}=\delta_{\sigma\sigma'}\sum_{\sigma''}
\int\frac{dE}{2\pi}~
(-i)G^<_{\sigma''\sigma''}(E)V_{\sigma''\sigma}
\label{hartree}
\eeq
and $\Sigma^r$ ($\Sigma^<$) is the retarded (lesser) impurity
self-energy, describing the effects of electron-correlation inside 
the junction.
The electron occupation numbers appearing in Eq. \eqref{Gless} are the usual statistical factors for a system of electrons:
$f_{L(R)}(\omega)=1/\{exp[(\omega-\mu_{L(R)})/k_BT]+1\}$.
 Since we operate in the regime of very small Kondo temperature,
we envision choosing 
an experimentally relevant
temperature that is large compared to $T_K$,
but which is much smaller than the coupling to the electrodes.

The other two non-equilibrium impurity Green's functions can be
simply obtained using:
\beq
G^a(\omega)=G^r(\omega)^{\dagger} 
\eeq
\beq
G^>(\omega)=G^r(\omega)-G^a(\omega)+G^<(\omega)
\eeq
\subsection{The $GW$ approximation for the impurity self-energy}
In the $GW$ approximation for the electron
self-energy, one does perturbation theory in terms of the screened
interaction $W$, keeping the first term in the expansion, the so
called $GW$ diagram. The $GW$ approximation has long been successfully used
in describing the equilibrium quasiparticle properties of real materials 
\cite{Hybertsen,Aryasetiawan,Aulbur00,Stan06,Schilfgaarde06}. It has
also been applied to the study
of real materials out of equilibrium, such as highly irradiated
semiconductors \cite{Spataru}, or, more recently, in transport
calculations through molecular nanojunctions \cite{Thygesen,Darancet,Thygesen08}.
For equilibrium properties the $GW$ approximation has been compared to a numerically exact
quantum Monte Carlo treatment \cite{Wang}; 
it has been found to be
adequate for small interactions or for high T, but not in the mixed valence or Kondo regimes.

Within the out-of-equilibrium $GW$ approximation, the general self-energy
expressions have the following form in frequency space \cite{Spataru}:
\beq
\Sigma^r_{\sigma\sigma'}(\omega)=i\int\frac{dE}{2\pi}~G^<_{\sigma\sigma'}(E)W^r_{\sigma\sigma'}(\omega-E)+i\int\frac{dE}{2\pi}~G^r_{\sigma\sigma'}(E)W^>_{\sigma\sigma'}(\omega-E)
\eeq
\beq
\Sigma^<_{\sigma\sigma'}(\omega)=i\int\frac{dE}{2\pi}~G^<_{\sigma\sigma'}(E)W^<_{\sigma\sigma'}(\omega-E)
\eeq
where the screened interaction $W$ can be obtained from the irreducible
polarizability $P$ through:
\beq
W^r(\omega)=[I-VP^r(\omega)]^{-1}V
\eeq
\beq
W^<(\omega)=W^r(\omega)P^<(\omega)W^a(\omega)
\eeq
\beq
W^>(\omega)=W^r(\omega)P^>(\omega)W^a(\omega)
\eeq

The irreducible polarization $P$ is evaluated in the random phase 
approximation (RPA):
\beq
P^r_{\sigma\sigma'}(\omega)=-i\int\frac{dE}{2\pi}~G^r_{\sigma\sigma'}(E)~G^<_{\sigma'\sigma}(E-\omega)-i\int\frac{dE}{2\pi}~G^<_{\sigma\sigma'}(E)~G^a_{\sigma'\sigma}(E-\omega)
\eeq
\beq
P^a_{\sigma\sigma'}(\omega)=-i\int\frac{dE}{2\pi}~G^a_{\sigma\sigma'}(E)~G^<_{\sigma'\sigma}(E-\omega)-i\int\frac{dE}{2\pi}~G^<_{\sigma\sigma'}(E)~G^r_{\sigma'\sigma}(E-\omega)
\eeq
\beq
P^<_{\sigma\sigma'}(\omega)=-i\int\frac{dE}{2\pi}~G^<_{\sigma\sigma'}(E)~G^>_{\sigma'\sigma}(E-\omega)
\eeq
Setting $P=0$ yields the Hartree-Fock approximation.

The set of equations for $G$, $\Sigma$, $W$ and $P$ are solved to
self-consistency, starting from an initial condition for $G$. All the
quantities are calculated on a real frequency grid 
(either regular or log-scale),
with an $\omega$-range up to $\pm 10 t$. 
Real and imaginary parts of the various
quantities are calculated explicitly, making sure that the retarded
functions obey the Kramers-Kronig relation.
In order to speed up the self-consistent process, we employ the Pulay scheme to
mix the Green's functions using previous iterations solutions
\cite{Pulay,Thygesen07b}:
\beq
{\cal G}^{j+1}_{in}=(1-\alpha)\bar{{\cal G}}^j_{in}+\alpha \bar{{\cal G}}^j_{out}
\eeq
where $\bar{{\cal G}}^n$ are constructed from the previous $m$ iterations:
\beq
\bar{{\cal G}}^j=\sum_{i=1}^m \beta_i {\cal G}^{j-m+i}
\eeq
and we choose three components for the parameter vector ${\cal G}$: $\Re G^r$, $\Im
G^r$ and $\Im G^<$. The values of $\beta_i$ are obtained by minimizing the distance
between $\bar{{\cal G}}_{in}^j$ and $\bar{{\cal G}}_{out}^j$. The scalar
product in the parameter space is defined using the integral in
Fourier space of a product of the component Green's
functions. We
found the speed of the convergence process to be quite independent on
the choice of reasonable values for 
$m$, as well as on the number of components for the parameter vector ${\cal G}$. As for the parameter $\alpha$, smaller values ($<0.1$) 
were needed for small bias voltages ($V<0.5$), while $\alpha=0.4$ was sufficient in order to achieve fast convergence for larger biases.

\subsection{Relation to physical observables}
The Green's functions of the impurity can be used to extract information
about observables pertaining to the impurity or even to the leads.
Thus, the spectral function of the impurity $A(\omega)$ is simply
related to the retarded Green's function:
\beq
A(\omega)=-\frac{1}{\pi}~Tr {\Im}G^r(\omega)
\eeq
where Tr stands for trace over the impurity spin degrees of freedom.
Also, the average impurity spin occupation number is: 
\beq
\langle n_{0,\sigma}\rangle=\int\frac{d\omega}{2\pi i}~G^<_{\sigma\sigma}(\omega)
\eeq
The expression for the average current passing through the junction is given by the
general Meir-Wingreen expression \cite{Meir}, which can be recast as
(see Appendix A for the derivation):
\beq
I=\int d\omega
~[f_L(\omega)-f_R(\omega)]~Tr\{\Gamma_L(\omega)~G^r(\omega)~\Gamma_R(\omega)~G^a(\omega)\} 
\nonumber
\eeq
\beq
+\int d\omega ~Tr\{[\Gamma_L(\omega)-\Gamma_R(\omega)]~G^r(\omega)~[\frac{i}{2}\Sigma^<(\omega)]~G^a(\omega)\}
\nonumber
\eeq
\beq
+\int d\omega
~Tr\{[f_L(\omega)\Gamma_L(\omega)-f_R(\omega)\Gamma_R(\omega)]~G^r(\omega)~
[-{\Im}\Sigma^r(\omega)]~G^a(\omega)\}
\label{Current}
\eeq
The first (Landauer type) term plays an important role whenever
correlations beyond the Hartree-Fock level are not considerable. It
gives the coherent component of the current.
 The second term is in
general very small for symmetric leads with relatively wide bands,
when $\Gamma_L(\omega)\approx\Gamma_R(\omega)$. The last term becomes important
when the electron-electron correlation effects are such that $-{\Im}\Sigma^r \approx \Gamma_{L(R)}$.

Having an expression for the average energy associated with the
junction for non-equilibrium can be
useful for a number of purposes, including calculation of current 
dependent forces \cite{diVentra}.
By formulating this as
the difference ${\cal \delta E}$ between the average energy of 
the total system (leads coupled to impurity) and the average energy of the isolated
leads, a finite result can be obtained. 
This can be done starting with
the following expression for the total average energy of the
system \cite{Kadanoff}:
\beq
{\cal E}=\frac{1}{2}\int\frac{d\omega}{2\pi i}~{\tilde{Tr}}\{(H_0+\omega I)~G^<(\omega)\}
\eeq 
where the trace ${\tilde{Tr}}$ is taken over a complete set of states spanning the
junction (indices n) and the leads (indices $k$). 
Alternatively, an equation of motion approach can be used \cite{KSS}.
We find that the two approaches give the same results.
The first approach is presented in Appendix B. 
Naturally, the energy
can be decomposed into
three terms, related respectively to the average
energy of the impurity ${\cal E}_{imp}$, the average energy of interaction
between leads and impurity ${\cal E}_{imp-leads}$, and the average energy
difference in the leads before and after adding the impurity $\delta{\cal E}_{leads}$:
\beq
{\cal \delta E}={\cal E}_{imp}+{\cal E}_{imp-leads}+\delta{\cal E}_{leads}
\label{Energy_first}
\eeq
where:
\beq
{\cal E}_{imp}=\frac{1}{2}\int\frac{d\omega}{2\pi i}~(\omega+V_G)~TrG^<(\omega)
\eeq
\beq
{\cal E}_{imp-leads}=\int\frac{d\omega}{2\pi i}~Tr\{[\Re\Delta_L^r(\omega)+\Re\Delta_R^r(\omega)]~G^<(\omega)~~~~~~~~~~~~~~~~~~~~~~~~~~
\nonumber
\eeq
\beq
~~~~~~~~~~~~~~~~~~~~~~~~~~-i[f_L(\omega)~\Im\Delta^r_L(\omega)+f_R(\omega)~\Im\Delta^r_R(\omega)]~[G^a(\omega)+G^r(\omega)]\} 
\label{Energy_imp_leads}
\eeq
\beq
\delta {\cal E}_{leads}=\frac{1}{2}\int\frac{d\omega}{2\pi  i}
~Tr\{[\Re F_L(\omega)+\Re F_R(\omega)]~G^<(\omega)~~~~~~~~~~~~~~~~~~~~~~~~~~
\nonumber
\eeq
\beq
~~~~~~~~~~~~~~~~~~~~~~~~~~-i[~f_L(\omega)~\Im F_L(\omega)+f_R(\omega)~\Im F_R(\omega)]~[G^a(\omega)+G^r(\omega)]\}
\label{Energy_leads}
\eeq
with:
\beq
F_{L(R)_{nm}}(\omega)=
-\Delta^r_{L(R)}(\omega)-2\omega\frac{d}{d\omega}\Delta^r_{L(R)}
\label{func_F}
\eeq
We note that the average energy change in the two leads is
always finite in the steady state case. 
A similar statement holds for 
the average number of electrons
displaced in the two leads $\delta N_{leads}$ (explicit expression in Appendix C).

\section{Results}
\subsection{Coulomb blockade}
\begin{figure}
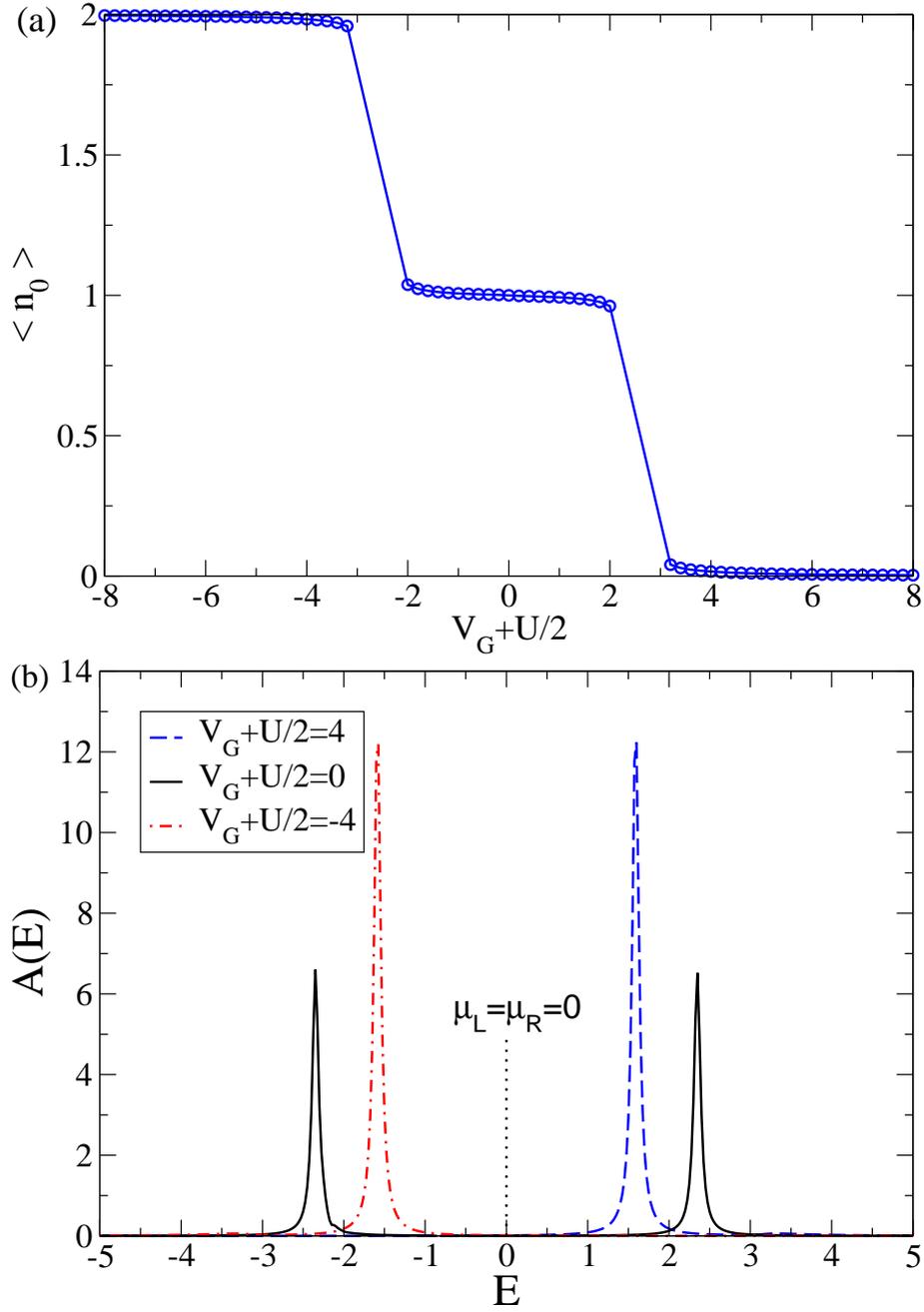

\resizebox{12.0cm}{!}{\includegraphics[clip=true]{fig2a.eps}}
\resizebox{12.0cm}{!}{\includegraphics[clip=true]{fig2b.eps}}
\caption{
Results for the self-consistent $GW$ approximation at zero applied source-drain bias and $k_BT=0$. 
(a) Impurity occuption number as a function of gate voltage.
(b) Spectral function for three different values of the gate voltage
$V_G$. Using $U=4.78$ and $\Gamma=0.05$.}
\label{staircase}
\end{figure}

In the weak coupling/strong interaction regime, the electron transport through
a junction can be blocked due to the charging energy in the junction.
Figure \ref{staircase}(a) shows the calculated impurity occupation 
number $\langle n_0 \rangle=\langle n_{0 \uparrow}\rangle+\langle n_{0
  \downarrow}\rangle$ as a function of the gate voltage $V_G$, 
at zero applied bias $V=0$ 
\cite{T_used}.
One can clearly see the Coulomb
staircase. The electron-hole symmetry of the 
Hamiltonian describing the system,~$H$, insures that the spectral 
function satisfies: $A(\omega;V_G+U/2)=A(-\omega;-V_G-U/2)$. As a
consequence, one has:
$\langle n_0(V_G+U/2) \rangle=2-\langle n_0(-V_G-U/2) \rangle$.  
A similar Coulomb staircase picture can be obtained at the
Hartree-Fock approximation level.

The impurity occupation number evolves from $0$ to $2$ as $V_G$ is
decreased from positive to negative values. 
Figure \ref{staircase}(b) shows the evolution of the spectral function
for three representative values of $V_G$. 
For $V_G+U/2=\pm4$, the solution is non-magnetic, 
with both spin levels degenerate, empty or occupied.
At the symmetric point (half-filling) $V_G+U/2$=0, the solution is a 
broken symmetry magnetic ground-state, with one spin occupied and the other
empty. 
Since we consider  temperatures that, although small, are still large compared to $T_K$,
the degenerate magnetic ground state is an appropriate representation of the physics.
In Fig. \ref{staircase}(a), the magnetic solution is found for
$|V_G+U/2|<2$; for $2<|V_G+U/2|<3$, a well converged (non-magnetic)
solution could not be found at $k_BT=0$.

\begin{figure}
\resizebox{12.0cm}{!}{\includegraphics[clip=true]{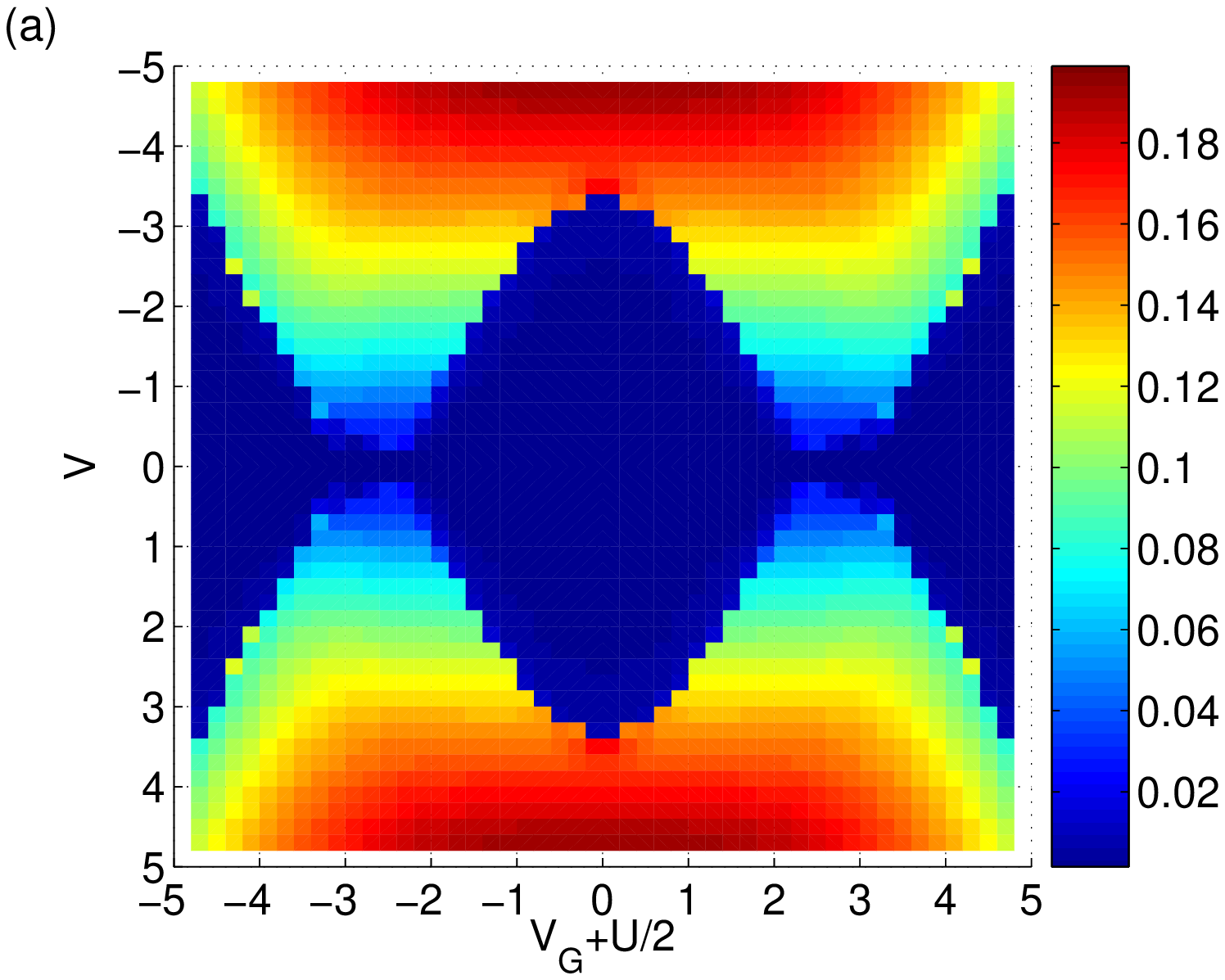}}
\resizebox{12.0cm}{!}{\includegraphics[clip=true]{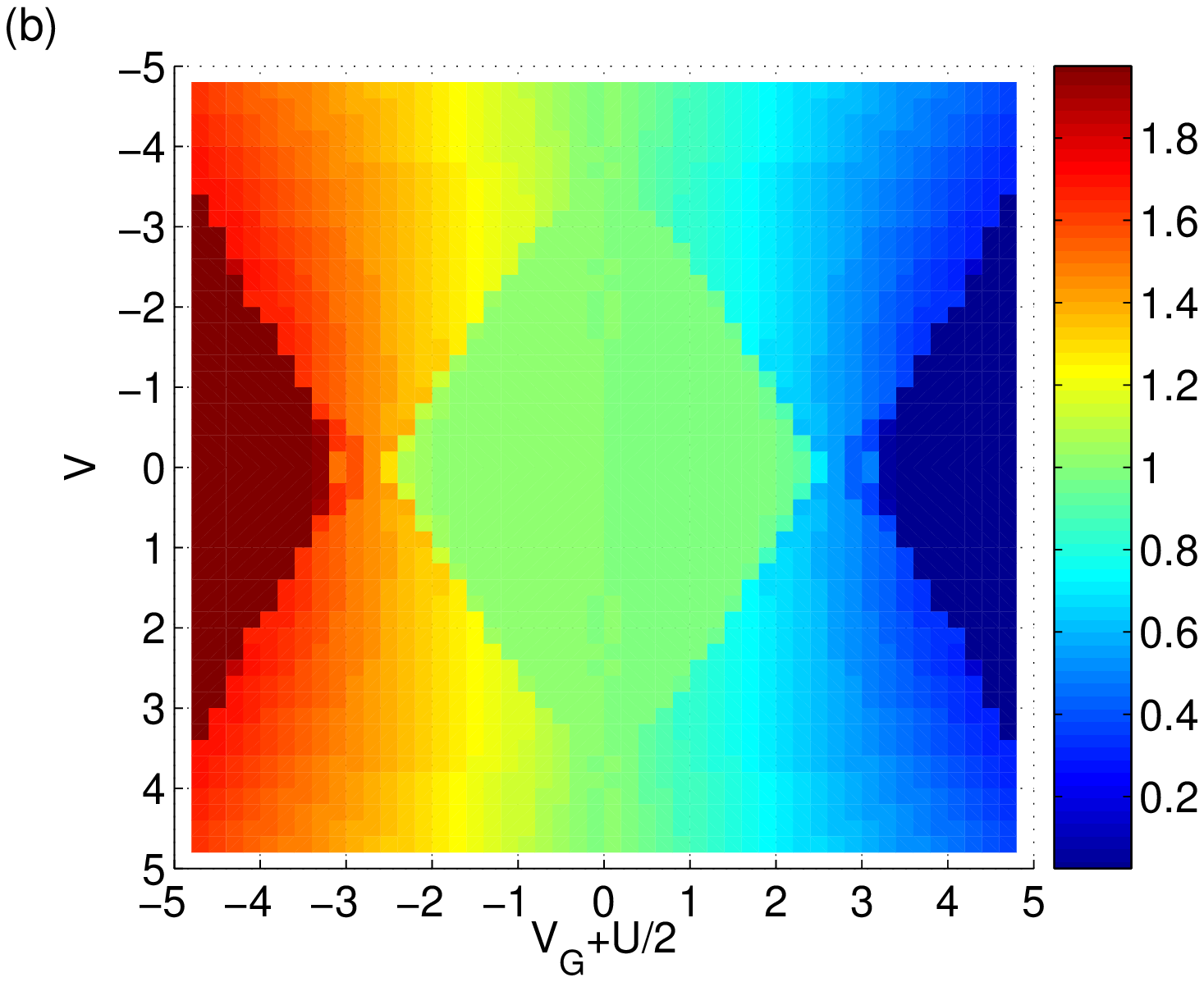}}
\caption{
False color plots of junction properties 
calculated in the self-consistent $GW$ approximation
as a function of the applied source-drain bias $V$ and gate voltage
$V_G$ at $k_BT=0$.
(a) Current.
(b) Average impurity occupation number. Using $U=4.78$ and $\Gamma=0.05$.}
\label{diamonds}
\end{figure}

Figure \ref{diamonds}(a) shows a color-scale plot of the current
$I$ as a function of the applied bias $V$ and gate voltage $V_G$. 
The plot is obtained by forward scan of the bias, i.e. using the lower 
bias solution as starting input for the higher bias calculation. One can
see the formation of Coulomb diamonds, inside which the current is
negligible, a signature of the Coulomb blockade
regime. A similar color-scale plot of the differential conductivity would show
sharp peaks at the edges of the Coulomb diamonds, but no tunneling
channel in the zero bias region inside the central Coulomb
diamond. Such a tunneling channel is absent in experiments on 
single-molecule transistors characterized by weak coupling between
molecules and electrodes \cite{McEuen}, but  has been 
observed when coupling to the electrodes is strong enough that
the Kondo temperature is appreciable $T_K\sim 10$ to 30 meV \cite{McEuen,Park}. 

At zero bias, zero temperature and at the symmetric point, the unitarity limit
\cite{Abrikosov} requires that the differential conductivity equals
$2e^2/h$. 
The broken (magnetic) symmetry solution in the $GW$ approximation
in the strong interaction regime does not
satisfy the unitarity limit; the spectral function does not have
the correct height near the chemical potential. 
Therefore, the
$GW$ approximation can not account
for the zero-bias tunneling channel observed for $T < T_K$. 
Under finite bias, the differential conductance due to the Kondo peaks in the
spectral function must fall off once the bias exceeds the Kondo
temperature \cite{MeirWingreenLee}; the Kondo peak
splits under non-zero bias, following the two different chemical
potentials and broadens quickly with increasing bias. 
Therefore, the width in applied bias for which such a
channel would be observed in the exact theory is of order $T_K$.
For the strong interaction regime considered here, this is negligible.
Thus, the $GW$ approximation provides the correct qualitative features of the Coulomb
blockade regime, namely Coulomb diamonds with no Kondo-assisted
conductance channels. 

The size of the Coulomb diamond depends on the interplay between the
repulsion U and the coupling to the leads, $\Gamma$. In the limit of
$U/\Gamma\rightarrow\infty$ the system becomes effectively an isolated
ion, and the size of the diamond is set by U. 
In our case,
$U/\Gamma\approx 100$ and the computed size of the Coulomb diamond
is only slightly smaller (by $\approx 20\%$) in the $GW$ approximation than 
in the Hartree-Fock approximation.
However, we suspect that the magnetic solution found in the $GW$ 
approximation underestimates the electronic
correlation originating from spin-spin quantum fluctuations,
and thus a more exact theory should result in smaller
size Coulomb diamonds than the ones we find.

The corresponding average electron occupation number 
$\langle n_0 \rangle$ is shown in Fig. \ref{diamonds}(b), where we can 
see that $\langle n_0 \rangle$ takes integer values of $0$, $1$ and
$2$ inside the Coulomb diamonds. For a given gate voltage, the spectral
function of the system changes appreciably only when the left or right lead
Fermi levels get closer to one of the impurity resonance
levels. As soon as a resonant level is pinned by a Fermi level, the
current increases while the impurity occupation number either
increases or decreases depending whether the pinned level is empty or 
occupied.

\subsection{Hysteresis in the I-V characteristics}
In an earlier study \cite{Wang} we concluded that, in the regime of 
intermediate strength of the Coulomb interaction, the $GW$ 
approximation leads to a broken spin symmetry ground state and thus 
fails to describe the spectral function correctly, missing 
completely the Kondo peak. 
A non-magnetic solution in the interaction regime
 $U/\Gamma>8$ has been elusive for other authors as well \cite{Thygesen07b}.
Recently, by employing a logarithmic frequency scale 
near the Fermi level, we have been able to find a non-magnetic 
solution in the strong interaction regime up to $U/\Gamma\approx25$ and $k_BT=0$.
Our results \cite{Spataru_tobepublished} show that equilibrium properties of 
the Anderson model, such as the total energy, Kondo temperature, T-linear coefficient 
of the specific heat or linear response conductance, are not satisfactorily 
described by the non-magnetic solution in the $GW$ approximation, 
as it was previously noted for several of these 
properties \cite{Wang,White}.

\begin{figure}
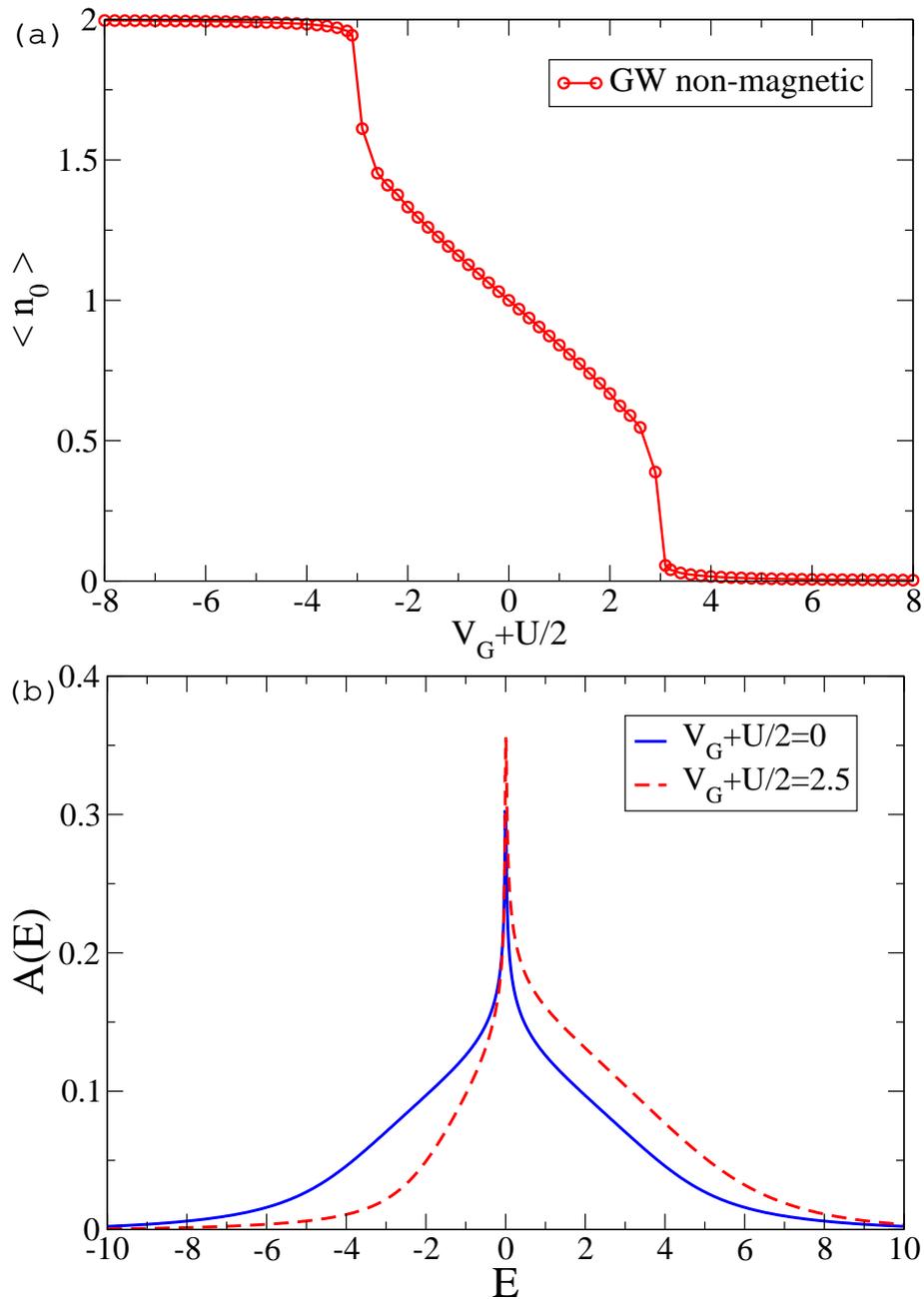

\resizebox{12.0cm}{!}{\includegraphics[clip=true]{fig4a.eps}}
\resizebox{12.0cm}{!}{\includegraphics[clip=true]{fig4b.eps}}
\caption{
Results for a non-magnetic solution throughout the gate bias range
in the $GW$ approximation at zero source-drain bias and $k_BT=0.01$. 
(a) Impurity occuption number as a function of gate voltage. 
(b) Spectral function for two values of gate voltage, the symmetric
case and an asymmetric case. Using $U=4.78$ and $\Gamma=0.05$.}
\label{nocc_nonmag}
\end{figure}

For the interaction strength considered in the present work, 
$U/\Gamma\approx100$, 
we have been able to calculate the non-magnetic solution at zero bias 
by considering small non-zero temperatures. 
We will consider $k_BT=0.01$ 
throughout the rest of the paper.
Figure \ref{nocc_nonmag}(a) 
shows the impurity occupation number as a function of gate voltage
for the non-magnetic solution. We see that
the Coulomb blockade plateau is not properly described; 
the impurity occupation 
number changes linearly about the symmetric point $V_G+U/2=0$. 
Figure \ref{nocc_nonmag}(b) 
shows the spectral function associated with the non-magnetic 
solution for two representative cases. 
In the symmetric case, one sees a broad peak 
(whose width is set by $U$) with a 
narrow portion near $E=0$ (whose width is set by $k_BT$). 
As the gate voltage is changed from the 
symmetric point, the narrow portion remains pinned near $E=0$, 
but the broad peak shifts together 
with $V_G$, hence the linear change in $<n_{occ}>$ as observed 
in Fig. \ref{nocc_nonmag}(a). 
Near $V_G+U/2=\pm2.8$, the non-magnetic solution cannot 
sustain a narrow portion near $E=0$, and 
the solution jumps into a phase with one narrow peak 
(of width $\sim\Gamma$) away from $E=0$ (as seen in 
Fig. \ref{staircase}(b) for $V_G+U/2=4$). 
In the region of gate bias near the transition at $V_G+U/2=\pm2.8$, 
the calculations get increasingly difficult to converge; 
for some values of the gate voltage 
a converged solution with retarded functions 
obeying the Kramers-Kronig relation could not be found.

\begin{figure}
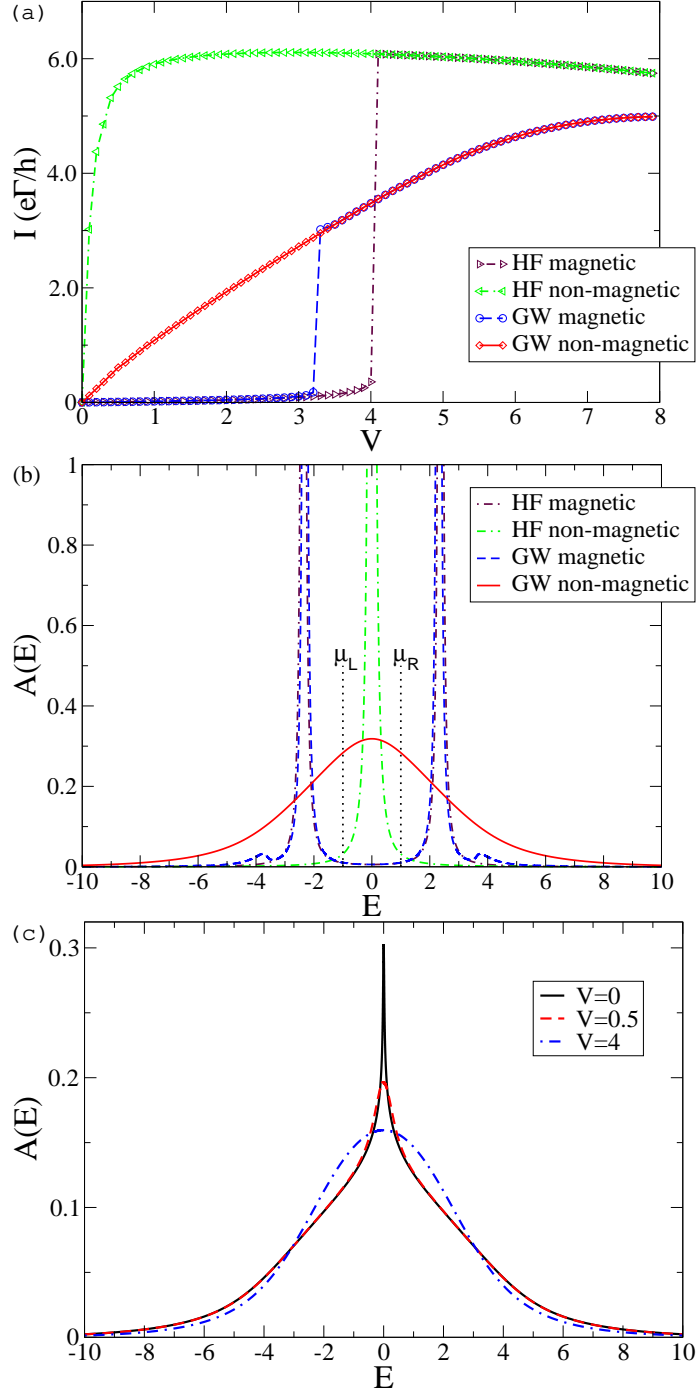

\resizebox{9.0cm}{!}{\includegraphics[clip=true]{fig5a.eps}}
\resizebox{9.0cm}{!}{\includegraphics[clip=true]{fig5b.eps}}
\resizebox{9.0cm}{!}{\includegraphics[clip=true]{fig5c.eps}}
\caption{
(a) Current as a function of the applied bias 
for gate voltage fixed to the symmetric case and $k_BT=0.01$.
Curves labeled magnetic correspond to a bias sweep from V=0 to V=8.
Curves labeled non-magnetic correspond to a reverse bias sweep from V=8 to V=0.
Results for the Hartree-Fock and $GW$ approximations are compared.
(b) Corresponding spectral functions for applied source-drain bias $V=2$. 
(c) Comparison of spectral functions for the non-magnetic solution in the $GW$ approximation
at three different applied source-drain bias values. Using $U=4.78$ and $\Gamma=0.05$.} 
\label{hysteresisI_A}
\end{figure}

Figure \ref{hysteresisI_A}(a) shows the current through the junction $I$ as a
function of the applied bias $V$ for a specific gate voltage $V_G$, 
$V_G+U/2=0$ such that the system is at half-filling, 
$\langle n_0 \rangle=1$. 
The general results do not depend on this symmetry.
The same qualitative results hold for a broad range of gate bias $|V_G+U/2|<2$.
Results obtained both in the $GW$ and Hartree-Fock approximation are shown.
These include a forward scan, starting from zero applied bias,
and a reverse scan starting from $V=8$.
At zero bias, 
we start with the magnetic solution, 
with one spin level occupied and the other one empty, as shown by the
solid-line curve of Fig. \ref{staircase}(b). 
Then in the forward scan,
the initial input at higher
bias is taken from the converged solution at lower bias.
For the reverse scan, the opposite approach is taken.
Note that the use of $k_BT=0.01$ 
has essentially no effect on the results except 
for the reverse scan with $V<0.5$ where the finite temperature
helps to stabilize the self consistent magnetic solution. 
Also, for reference, the $I-V$ data shown in Fig. \ref{diamonds}(a)
was obtained by forward bias scan. 

In both the Hartree-Fock and the $GW$ approximation, as the bias is increased, the
two spin levels remain outside the bias window and the current is
negligible until $V$ approaches a value of order (but less than) $U$.  
In Hartree-Fock this value is $V\approx 4.0$, while in $GW$ it is $V\approx 3.2$. 
At this point, where the broadened impurity levels get
pinned by the two chemical potentials, the character of the steady-state
solution changes from magnetic to non-magnetic.
At this bias, the current increases suddenly. 
Correspondingly, the spectral function
shows one double-degenerate peak 
centered half-way in between the two chemical potentials (Fig. \ref{hysteresisI_A}(b)). 
For higher bias, the Hartree-Fock and $GW$ approximations result in qualitatively
and quantitatively different behavior.
In Hartree-Fock, the current is approximately pinned 
at the value expected for a single, half-filled resonance in the bias
window ($2\pi\Gamma e/h$).
The overall downward drop is explained by the finite band width of the electrodes.
However, in the $GW$ approximation, the spectral function shows substantially larger
broadening and the current increases steadily with bias as the
spectral weight inside the bias window increases. 
Correspondingly, upon analysis of contributions to the current in this regime, it is largely due to
non-coherent transport, as $-{\Im}\Sigma^r>>\Gamma$ and
the main component of the current is given by the last term of 
the right hand side of Eq. \eqref{Current}.
The backward bias scan is started from the
non-magnetic solution at $V=8$. As the bias is decreased, the solution
remains non-magnetic well below the transition bias point from the forward bias scan, 
resulting in hysteresis in the $I-V$ curve.  
While in Hartree-Fock, the current remains high down to relatively low applied bias, 
the calculated current in the $GW$ approximation drops approximately linearly.

The physical description of the magnetic solution is straightforward.
The spectral function shows two peaks (Fig. \ref{hysteresisI_A}(b)),
spin up and spin down, one occupied and the other one empty, separated in
frequency by a little less than $U$. The results from the $GW$ approximation
are very close to those from Hartree-Fock approximation in this case.
There are very few 
occupied-to-empty electron-hole same-spin excitations; the polarization $P$ is very small. 

\begin{figure}
\resizebox{12.0cm}{!}{\includegraphics{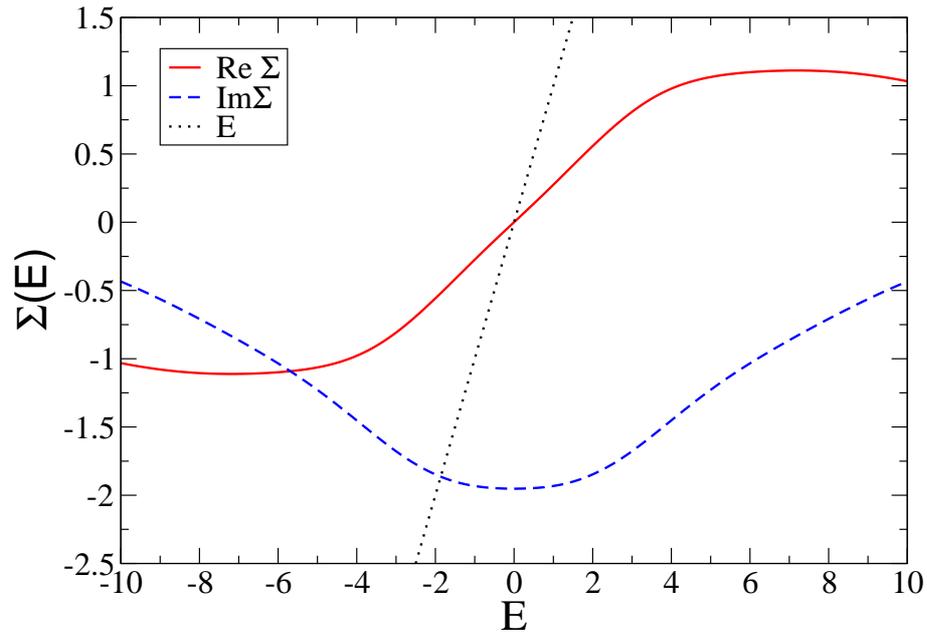}}
\caption{Real and imaginary parts of the retarded self-energy 
in the $GW$ approximation for the non-magnetic solution
at half-filling, applied source-drain bias $V=2$ and $k_BT=0.01$. 
Using $U=4.78$ and $\Gamma=0.05$.}
\label{Sigma_GW}
\end{figure}

The non-magnetic solution is more complex
and the physical picture is rather different for the
Hartree-Fock and the $GW$ approximations. 
While for Hartree-Fock 
the spectral function showing only one sharp peak
with width equal to $\Gamma$,
the spectral function in the $GW$ approximation is much broader (Fig. \ref{hysteresisI_A}(b)).
While the overall 
broadening depends strongly on the interaction parameter $U$, 
the applied bias $V$ affects the region of width $V$ about $E=0$ (Fig. \ref{hysteresisI_A}(c)).
Furthermore, the width of the spectral function 
is almost independent of the effective coupling coefficient $\Gamma$.
For example, for $k_BT=0.01$ and $V\in[0,8]$, 
the spectral function plot for $\Gamma=0.1$ is almost undistinguishable from that for the $\Gamma=0.05$ case.
This indicates that the broadening is due to quantum fluctuations taking place on the impurity.
The applied bias dependent broadening  
can be traced back to the large imaginary part of the 
retarded self-energy, as shown in Fig. \ref{Sigma_GW}. 
At zero bias 
and zero temperature
the Fermi liquid behavior of the system guarantees ${\Im}\Sigma^r(0)=0$. 
The non-zero value of ${\rm Im}\Sigma^r(0)$ 
shown in Fig. \ref{Sigma_GW}
is clearly a non-equilibrium, non-zero bias effect. 
A similar broadening, increasing strongly with bias, has been also observed 
in recent calculations based on the $GW$ approximation for a two-level model molecule \cite{Thygesen08}. 

\begin{figure}
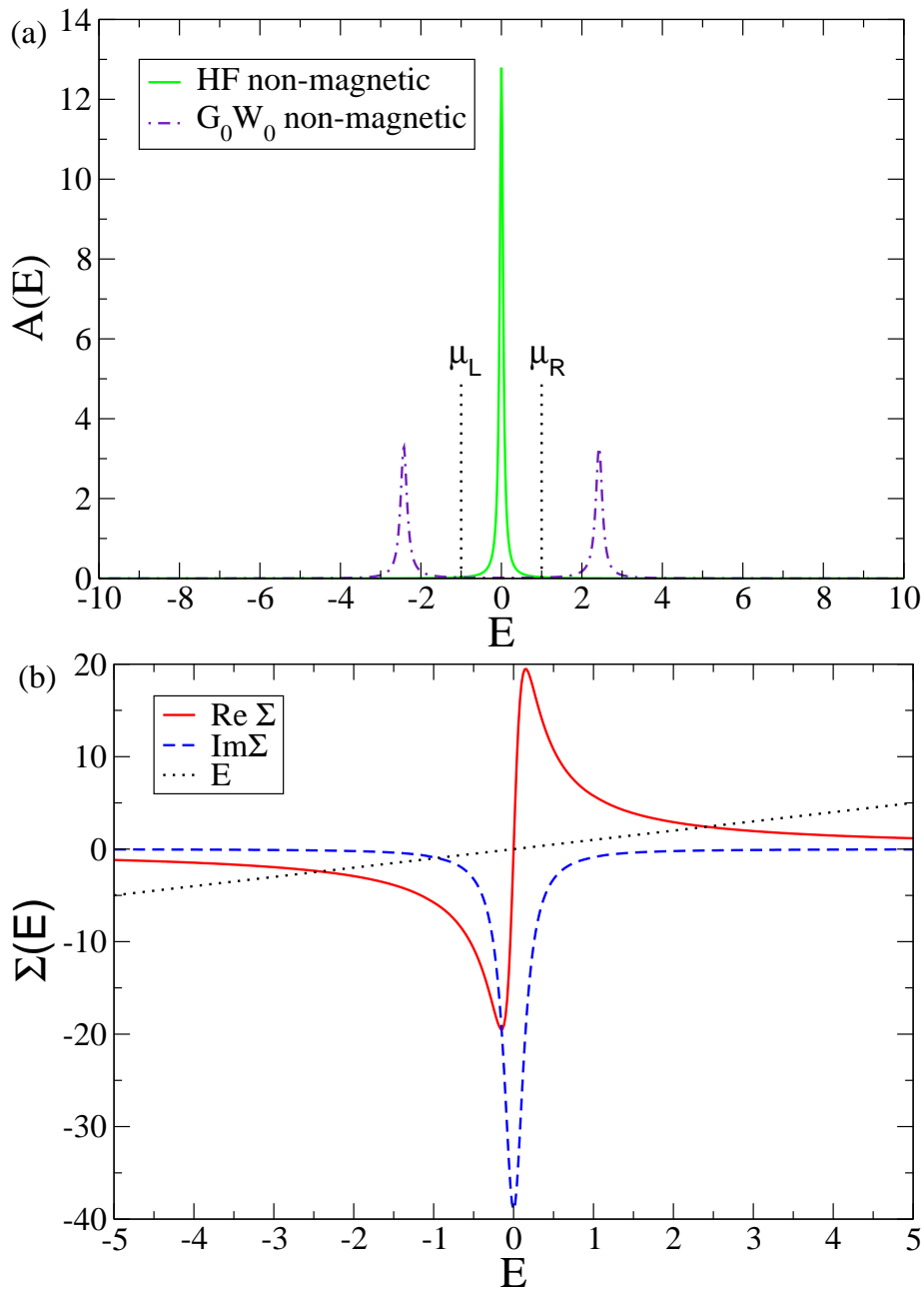

\resizebox{12.0cm}{!}{\includegraphics[clip=true]{fig7a.eps}}
\resizebox{12.0cm}{!}{\includegraphics[clip=true]{fig7b.eps}}
\caption{
Illustration of steps in the iterative solution
to arrive at the final non-magnetic solution in the $GW$ approximation.
Gate voltage fixed to the symmetric (half-filling) case, 
applied source-drain bias $V=2$ and $k_BT=0.01$.
(a) Spectral function for the non-magnetic Hartree-Fock and
 $G_0W_0$ solutions. 
(b) Real and imaginary parts of the 
 retarded non-magnetic $G_0W_0$ self-energy. Using $U=4.78$ and $\Gamma=0.05$.}
\label{A_HF_G0W0_Sigma_G0W0}
\end{figure}

The broadening of the spectral function for the non-magnetic solution
in the GW approximation can be understood by looking at how the spectral function and the
retarded self-energy changes as we iterate the non-magnetic solution from
Hartree-Fock to $GW$.
Here we denote with $G_0W_0$ the 
intermediate solution obtained with the Hartree-Fock Green's functions as input.
At the Hartree-Fock level, the non-magnetic solution has one narrow 
central peak, with half-width at half-maximum approximately given by 
$\Gamma=0.05$. The entire peak is situated inside the bias window, as
shown in Fig. \ref{A_HF_G0W0_Sigma_G0W0}(a). 
In that energy range one can find both occupied and
empty (more exactly half-occupied) quasi-states of the same spin.
Now, such a quasi-state can easily decay into another quasi-state with lower
or higher energy, by emitting or absorbing an electron-hole same-spin 
excitation with energy within the bias window range.
Thus, ${\Im}\Sigma^r$, which is proportional to the inverse lifetime of
the quasi-state, becomes very large at the the $G_0W_0$ level, as seen
in Fig. \ref{A_HF_G0W0_Sigma_G0W0}(b). 
From the $G_0W_0$ result for ${\Re}\Sigma^r$ (related to
${\Im}\Sigma^r$ 
through a Kramers-Kronig relation), 
it follows that the  $G_0W_0$
spectral function shows two double degenerate (spin up and spin
down) peaks, situated outside the bias window
(Fig. \ref{A_HF_G0W0_Sigma_G0W0}(a)).
If, at each of the next
iterative steps $i$, we would use as input only the Green's
functions from iteration $i-1$, the spectral function would oscillate 
between the two types (Hartree-Fock and $G_0W_0$) of
solution. However, by means of
the Pulay mixing scheme, we are able to achieve convergence rather
fast, with the self-consistent $GW$ solution looking somehow in between 
Hartree-Fock and $G_0W_0$, as seen in Fig. \ref{hysteresisI_A}(b).

\begin{figure}
\resizebox{12.0cm}{!}{\includegraphics{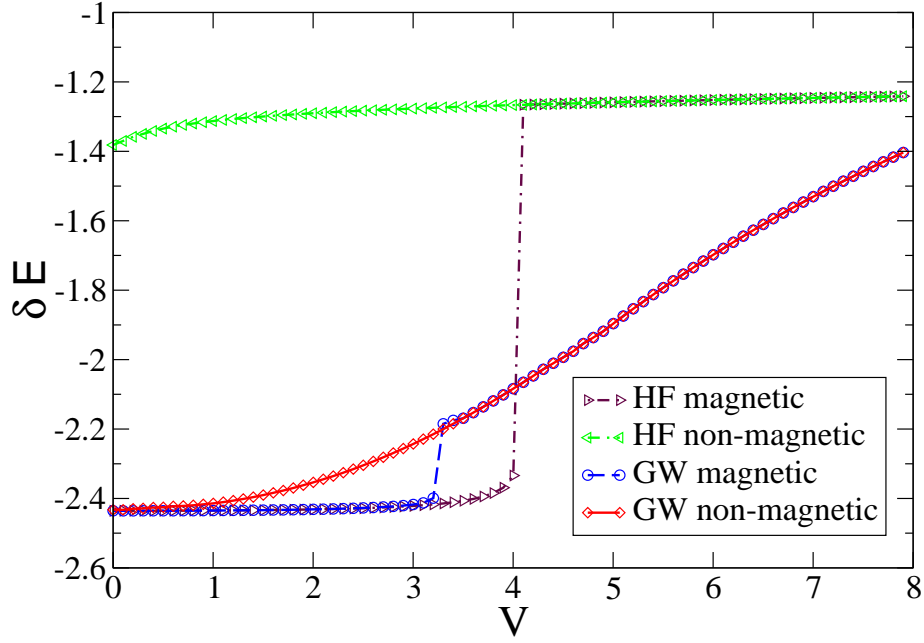}}
\caption{
Change in the average energy of the system 
as a function of applied source-drain bias 
for gate voltage fixed to the symmetric case (half-filling) and $k_BT=0.01$.
Curves labeled magnetic correspond to a bias sweep from V=0 to V=8.
Curves labeled non-magnetic correspond to a reverse bias sweep from V=8 to V=0.
Results for the Hartree-Fock and $GW$ approximations are compared.
 Using $U=4.78$ and $\Gamma=0.05$.}
\label{hysteresisE}
\end{figure}

The calculated I-V curves in Fig. \ref{hysteresisI_A}(a) result from 
the existence of two steady state solutions over a broad range of 
applied bias that are
accessed depending on initial conditions. Our procedure of stepping
the applied bias in forward followed by reverse scans with self
consistent solution at each step simulates an adiabatic voltage scan
and the existence of two stable solutions results in hysteresis.
One may ask whether quantum fluctuations that are beyond 
the scope of the $GW$ approximation would eliminate the hysteresis. 
To probe this, we need to understand the energy
difference between the system in the magnetic and the non-magnetic solutions
in the hysteretic region.
Figure \ref{hysteresisE} shows the change in the average energy of the 
total system, ${\cal \delta E}$, calculated as described in Section
III(B),
as a function of the applied bias at half-filling. 
Results are shown for both the Hartree-Fock and the $GW$ approximations,
following the same loop of forward and reverse bias scans.
For weak effective coupling between impurity and leads, for the magnetic solution, one has  ${\cal
  \delta E}\approx{\cal E}_{imp}+\textsl{O}(\Gamma)$. 
Near equilibrium, the magnetic solutions in the forward bias scan show very similar energies,
close to the energy
of the isolated, single-occupied impurity: 
${\cal \delta E}^{mag}\sim V_G=-U/2$.
However, at the applied bias where the current rapidly increases
and the solution changes to
non-magnetic, Hartree-Fock yields an average energy higher
than the magnetic one by about $U/4$.
On the other hand, the $GW$ approximation shows an average energy change that is much smaller.
Correspondingly, on the reverse bias scan, the bias dependence of the average energy
is also much different.
While the energy in the Hartree-Fock approximation remains high as the bias approaches zero,
the energy in the $GW$ approximation approaches a value that is only
higher than the zero bias magnetic state by about $\Gamma/20$.

We have found that in the strong interaction regime, 
there are two distinct self consistent solutions with the $GW$ approximation.
These lead to hysteresis in the calculated $I-V$ curves.
However, at zero bias, bistability is forbidden for 
the Anderson model \cite{Mitra05,Mitra07}. 
Therefore, the states represented by those solutions found in the $GW$ approximation
must be unstable with respect to
quantum fluctuations that have not been taken into account.
The fact that the average energy of 
the magnetic state is lower than that of
non-magnetic solution, is 
probably an indication of the larger weight of the 
magnetic solution in the emerging exact many-body state. 
As the bias is increased away from equilibrium, 
Fig. \ref{hysteresisE} shows that the 
energy difference between non-magnetic and magnetic
configurations also increases in the $GW$ approximation. 
However, for applied bias larger than about $\Gamma/20$, 
the energy difference 
is smaller than the applied bias.
This means that 
at non-zero biases
on-shell processes will be possible through which one configuration can decay 
into the other one
(with one electron transferring from one lead to the other to insure
total energy conservation). 
We thus expect that out-of-equilibrium, the lifetime of the $GW$ 
bistable states would be even smaller than at equilibrium.
Quantum fluctuations between the
two degenerate magnetic configurations and the non-magnetic one will 
eliminate the hysteresis and renormalize in a non-trivial way
the emerging unique many-body state. 
Therefore, the hysteresis in the $I-V$ curve is probably another signal that
the $GW$ approximation is not representing important aspects of the strong 
interaction regime. 
A calculation of the lifetime of the
bistable states found with the $GW$ approximation 
is beyond the scope of the present work, 
but would be very valuable.

\section{Summary}
In this work we used the GW approximation to study the role of 
electron-electron correlation effects in the out-of-equilibrium 
single impurity Anderson model. 
We considered the regime with weak level broadening and strong 
Coulomb interaction, treating the electron-electron interaction 
with the self-consistent $GW$ approximation for the electron self energy.  
We found that the GW approximation accounts 
for Coulomb blockade effects. 
The low conductance (blockade) region in gate bias and source-drain bias
corresponds to a magnetic solution in the GW approximation.
At the edge of the blockade region, the current jumps and
the self consistent solution changes to a non-magnetic character.
The position of the transition and the jump in current are 
renormalized from the Hartree-Fock values.
However, we also found a self consistent non-magnetic 
solution inside 
the Coulomb blockade region. 
As a consequence, the GW approximation also predicts an unphysical 
hysteresis in the I-V characteristics of the system. 
Outside the blockade region, e.g. where the source-drain bias is high
and the magnetic solution is not stable, we expect that the GW
approximation gives a reasonable account of the conductance.   
However, the jump in current at the edge of the blockade region 
and the hysteresis inside the blockade region both appear to arise 
from a first-order-transition-like bistability in the GW approximation.
An analysis of the total energy
difference between the magnetic and non-magnetic solutions
suggests that quantum fluctuations beyond the scope of the $GW$ 
approximation would result in rapid decay of the non-magnetic
solution, eliminating both the sharp jump and the hysteresis.

\section*{Acknowledgments}
This work was primarily supported by the Nanoscale Science and Engineering
Initiative of the National Science Foundation under
NSF Award Numbers CHE-0117752 and CHE-0641523, and by the New York
State Office of Science, Technology, and Academic Research (NYSTAR).
This work was partially supported by the National Science Foundation
under grant NSF-DMR-0705847 and grant NSF-DMR-0705941, and by the 
U. S. Department of Energy under Contract No. DE-AC01-94AL85000, 
Contract No. DE-AC02-98CH10886 and Contract No. DE-AC02-05CH11231. 
Sandia is a multiprogram laboratory operated by Sandia Corporation, 
a Lockheed Martin Company, for the United States Department of Energy.

\appendix
\renewcommand{\theequation}{A-\arabic{equation}}
\setcounter{equation}{0}
\section*{Appendix A: Average current through the junction}
We start from the Meir-Wingreen expression for the current from the
left lead (in units of $e=h=1$) \cite{Meir}:
\beq
I_L=i \int d\omega
~Tr\{\Gamma_L(\omega)~G^<(\omega)+f_L(\omega)\Gamma_L(\omega)~[G^r(\omega)-G^a(\omega)]\}
\equiv \int d\omega J_L(\omega).
\eeq
Using that in steady state $I=I_L=(I_L-I_R)/2$, and making use of Eq. \eqref{Gless} and the relation:
\beq
G^r(\omega)-G^a(\omega)=G^r(\omega)~[\Delta^r_L(\omega)+\Delta^r_R(\omega)+\Sigma^r(\omega)-h.c.]~G^a(\omega)
\eeq
one obtains:
\beq
I=\frac{1}{2}\int d\omega
~[f_L(\omega)-f_R(\omega)]~Tr\{\Gamma_L(\omega)~G^r(\omega)~\Gamma_R(\omega)~G^a(\omega)\} 
\nonumber
\eeq
\beq
+\frac{1}{2}\int d\omega
~[f_L(\omega)-f_R(\omega)]~Tr\{\Gamma_R(\omega)~G^r(\omega)~\Gamma_L(\omega)~G^a(\omega)\} 
\nonumber
\eeq
\beq
+\frac{i}{2}\int d\omega ~Tr\{[\Gamma_L(\omega)-\Gamma_R(\omega)]~G^r(\omega)~\Sigma^<(\omega)~G^a(\omega)\}
\nonumber
\eeq
\beq
+\frac{i}{2}\int d\omega ~Tr\{[f_L(\omega)\Gamma_L(\omega)-f_R(\omega)\Gamma_R(\omega)]~G^r(\omega)~[\Sigma^r(\omega)-\Sigma^r(\omega)^\dagger]~G^a(\omega)\}
\label{I}
\eeq
In the single impurity Anderson model case, the Green's functions are symmetric (the
off-diagonal elements being simply zero) and the first two terms in
Eq. \eqref{I} are equal, with the final expression for the current
reading as in Eq. \eqref{Current}.

\appendix
\renewcommand{\theequation}{B-\arabic{equation}}
\setcounter{equation}{0}
\section*{Appendix B: The change in energy caused by impurity }
For simplicity, we consider eigenstates of the non-interacting 
isolated junction (energies $\epsilon_n$) and
isolated leads (energies $\epsilon_k$). 
Denoting with $g$ the Green's function of the isolated lead, the 
difference between the average energy of the total system and the
average energy
of the isolated leads can be written:
\beq
{\cal \delta E}={\cal E}_{imp}+{\cal E}_{imp-leads}+\delta{\cal E}_{leads}
\label{Energy_first}
\eeq
where:
\beq
{\cal E}_{imp}
=\frac{1}{2} \sum_n \int\frac{d\omega}{2\pi i}~(\omega+\epsilon_n)
~G^<_{nn}(\omega)
\eeq
\beq
{\cal E}_{imp-leads}=\Re \sum_{n,k} \int\frac{d\omega}{2\pi  i}~H_{0,nk}~G^<_{k n}
\eeq
(we made use of the fact that $G^<(\omega)^\dagger=-G^<(\omega)$), and:
\beq
\delta{\cal E}_{leads}=
\frac{1}{2} \sum_{k} \int\frac{d\omega}{2\pi i}~(\epsilon_k+\omega)
~[G^<_{k k}(\omega)-g^<_{k k}(\omega)]
\eeq

The expression for $G^<_{k n}(\omega)$ can be derived rather
easily in the present case of non-interacting leads \cite{Jauho}:
\beq
G^<_{k n}(\omega)=\sum_m ~g^r_{k k}(\omega)~H_{0,k m}~G^<_{mn}(\omega)+
\sum_m ~g^<_{k k}(\omega)~H_{0,k m}~G^a_{mn}(\omega)
\label{G_alpha_n}
\eeq

Using:
\beq
\sum_k H_{0,nk}~g^r_{k k}(\omega)~H_{0,k
  m}=\Delta_L^r{_{nm}}(\omega)+\Delta_R^r{_{nm}} (\omega)
\eeq
and
\beq
\sum_k H_{0,nk}~g^<_{k k}(\omega)~H_{0,k
  m}=if_L(\omega)\Gamma_L{_{nm}}(\omega)+if_R(\omega)\Gamma_R{_{nm}}(\omega)
\eeq
one arrives at the following expression for ${\cal E}_{imp-leads}$:
\beq
{\cal E}_{imp-leads}=\Re \int\frac{d\omega}{2\pi i}~Tr\{[(\Delta_L^r(\omega)+\Delta_R^r(\omega)]~G^<(\omega)
+i[f_L(\omega)\Gamma_L(\omega)+f_R(\omega)\Gamma_R(\omega)]~G^a(\omega)\} 
\label{E_imp_leads_real}
\eeq
Using also the fact that the flux of particles coming in and out
from the junction is exactly zero in steady states:
\beq
\int d\omega ~[J_L(\omega)+J_R(\omega)]=0
\eeq
\beq
\Rightarrow
\int d\omega
~Tr\{[\Gamma_L(\omega)+\Gamma_R(\omega)]~G^<(\omega)-[f_L(\omega)\Gamma_L(\omega)+f_R(\omega)\Gamma_R(\omega)]~[G^a(\omega)-G^r(\omega)]\}
=0
\eeq
one can ignore taking the real part of the r.h.s. of Eq. 
\eqref{E_imp_leads_real}:
\beq
{\cal E}_{imp-leads}=\int\frac{d\omega}{2\pi i}~Tr\{[(\Delta_L^r(\omega)+\Delta_R^r(\omega)]~G^<(\omega)
+i[f_L(\omega)\Gamma_L(\omega)+f_R(\omega)\Gamma_R(\omega)]~G^a(\omega)\} 
\eeq
which can be further written as in Eq. \eqref{Energy_imp_leads}.

Now let's focus on the expression for $\delta{\cal E}_{leads}$.
Similarly to Eq. \eqref{G_alpha_n} one also has:
\beq
G^<_{k k}(\omega)=g^<_{k k}(\omega)+
\sum_n ~g^r_{k k}(\omega)~H_{0,k n}~G^<_{n k}(\omega)+
\sum_n ~g^<_{k k}(\omega)~H_{0,k n}~G^a_{n k}(\omega)
\eeq

Further use of:
\beq
G^<_{n k}(\omega)=\sum_m ~G^<_{n m}(\omega)~H_{0,m k}~g^a_{k k}(\omega)+
\sum_m ~G^r_{n m}(\omega)~H_{0,m k}~g^<_{k k}(\omega),
\label{G_n_alpha}
\eeq
\beq
G^a_{n k}(\omega)=\sum_m ~G^a_{nm}(\omega)~H_{0,mk}~g^a_{k k}(\omega),
\eeq
\beq
g^<_{k k_{L(R)}}(\omega)=f_{L(R)}(\omega)~[g^a_{kk_{L(R)}}(\omega)-g^r_{kk_{L(R)}}(\omega)],
\eeq
\beq
\sum_{k\in L(R)} (\epsilon_k+\omega)~H_{0,m k}~g^r_{k k}(\omega)~g^{
\begin{array}{ccc}
\hspace{-0.1cm} \vspace{-0.55cm} _a\\
\hspace{-0.1cm} \vspace{-0.1cm} _r
\end{array} 
}
_{k k}(\omega)~H_{0,kn}= \lim_{\delta\rightarrow 0}
\int \frac{d\epsilon }{2\pi}
\frac{(\epsilon+\omega)\Gamma_{L(R)_{mn}}(\epsilon)}{(\omega-\epsilon+i\delta)(\omega-\epsilon
  \mp i\delta)}
\eeq
allows us to write the expression for $\delta{\cal  E}_{leads}$
as:
\beq
\delta {\cal E}_{leads}=\frac{1}{2}\int\frac{d\omega}{2\pi  i}
~Tr\{[S_L(\omega)+S_R(\omega)]~G^<(\omega)-[S_L(\omega)~f_L(\omega)+S_R(\omega)~f_R(\omega)]~[G^a(\omega)-G^r(\omega)]\}
\nonumber
\eeq
\beq
-\frac{1}{2} [ \int\frac{d\omega}{2\pi  i} ~Tr\{
[F_L(\omega)~f_L(\omega)+F_R(\omega)~f_R(\omega)]~G^r(\omega)\}+h.c.]
\eeq
with:
\beq
S_{L(R)_{nm}}(\omega)=\lim_{\delta \rightarrow 0}\int
\frac{d\epsilon}{2\pi}
\frac{(\omega+\epsilon)~\Gamma_{L(R)_{nm}}(\epsilon)}{(\omega-\epsilon+i\delta)(\omega-\epsilon-i\delta)}
\eeq
\beq
F_{L(R)_{nm}}(\omega)=\lim_{\delta\rightarrow 0} \int
\frac{d\epsilon}{2\pi}
\frac{(\omega+\epsilon)~\Gamma_{L(R)_{nm}}(\epsilon)}{(\omega-\epsilon+i\delta)^2}
\eeq

The function $S(\omega)$ has a singular part which however doesn't 
contribute to $\delta{\cal E}_{leads}$. Indeed, writing:
\beq
S_{L(R)_{nm}}(\omega)=\Re
F_{L(R)_{nm}}(\omega)+\lim_{\delta\rightarrow 0}~\frac{1}{\pi}\int d\epsilon
~(\omega+\epsilon)~\Gamma_{L(R)_{nm}}(\epsilon)~\frac{\delta}{(\omega-\epsilon)^2+\delta^2}~\frac{\delta}{(\omega-\epsilon)^2+\delta^2},
\label{singular}
\eeq
the contribution to $\delta{\cal E}_{leads}$ of the second term on the r.h.s. of
Eq. \eqref{singular} is proportional to:
\beq 
\lim_{\delta\rightarrow 0} ~\frac{1}{\delta}\int d\omega~\omega
~Tr\{[\Gamma_L(\omega)+\Gamma_R(\omega)]~G^<(\omega)-[f_L(\omega)\Gamma_L(\omega)+f_R(\omega)\Gamma_R(\omega)]~[G^a(\omega)-G^r(\omega)]\}=0
\label{no_contrib}
\eeq
which vanishes by virtue of the fact that the integral multiplying $\frac{1}{\delta}$ is proportional to
the flux of energy coming in and out
from the junction, which is exactly zero in steady states:
\beq
\int d\omega ~\omega~[J_L(\omega)+J_R(\omega)]=0
\eeq

Thus, the expression for $\delta{\cal E}_{leads}$ becomes:
\beq
\delta {\cal E}_{leads}=\frac{1}{2}\int\frac{d\omega}{2\pi  i}
~Tr\{[\Re F_L(\omega)+\Re F_R(\omega)]~G^<(\omega)-[\Re F_L(\omega)~f_L(\omega)+\Re F_R(\omega)~f_R(\omega)]~[G^a(\omega)-G^r(\omega)]\}
\nonumber
\eeq
\beq
-\Re\int\frac{d\omega}{2\pi  i} ~Tr\{
[F_L(\omega)~f_L(\omega)+F_R(\omega)~f_R(\omega)]~G^r(\omega)\}
\label{app_en_leads}
\eeq
Noting that the function $F_{L(R)}(\omega)$ is related in a simple way to the energy
derivative of $\Delta^r_{L(R)}(\omega)$, one finally arrives at
Eqs. \eqref{Energy_leads}-\eqref{func_F}.

\appendix
\renewcommand{\theequation}{C-\arabic{equation}}
\setcounter{equation}{0}
\section*{Appendix C: Average number of displaced electrons in the leads}
In a manner similar to the one described in detail in Appendix B, one
can obtain an expression for the average number of electrons 
displaced in the two leads:
\beq
\delta N_{leads}\equiv \sum_{k} \int\frac{d\omega}{2\pi i}~
~[G^<_{k k}(\omega)-g^<_{k k}(\omega)]
\eeq
with the final expression reading:
\beq
\delta N_{leads}=-\int\frac{d\omega}{2\pi  i}
~Tr\{[\frac{d}{d\omega} \Re\Delta^r_L(\omega)+\frac{d}{d\omega} \Re\Delta^r_R(\omega)]~G^<(\omega)
\nonumber
\eeq
\beq
-i[~f_L(\omega)~\frac{d}{d\omega}\Im \Delta^r_L(\omega)+f_R(\omega)~\frac{d}{d\omega}\Im \Delta^r_R(\omega)]~[G^a(\omega)+G^r(\omega)]\}
\label{N_leads}
\eeq

\newpage

\clearpage
\newpage

\newpage

\clearpage
\newpage

\clearpage
\newpage

\clearpage
\newpage

\clearpage
\newpage

\clearpage
\newpage


\begin{references}

\bibitem{McEuen}
J. Park et al., Nature ${\bf 417}$, 722 (2002).

\bibitem{Park}
W. Liang et al., Nature ${\bf 417}$, 725 (2002).

\bibitem{Natelson}
D. Natelson, {\it Handbook of Organic Electronics and Photonics}
(American Scientific Publishers, 2006).

\bibitem{Burke}
C. Toher, A. Filippetti, S. Sanvito, and K. Burke, 
Phys. Rev. Lett. ${\bf 95}$, 146402 (2005).

\bibitem{Anders08} F. B. Anders, arXiv:0803.3004.

\bibitem{Weiss08} S. Weiss, J. Eckel, M. Thorwart, and R. Egger, Phys. Rev. B {\bf 77}, 195316 (2008).

\bibitem{Al-Hassanieh06} K. A. Al-Hassanieh, A. E. Feiguin, J. A. Riera, C. A. Busser and E. Dagotto, Phys. Rev. B {\bf 73}, 195304 (2006).
\bibitem{Kirino08} S. Kirino, T. Fujii, J. Zhao and K. Ueda, J. Phys. Soc. Jpn. {\bf 77}, 084704 (2008).

\bibitem{Muehlbacher08} L. M\"uhlbacher and E. Rabani, Phys. Rev. Lett. {\bf 100}, 176403 (2008).

\bibitem{Schiro08} M. Schiro and M. Fabrizio, arXiv:0808.0589.

\bibitem{Schmidt08} T. Schmidt, P. Werner, L. M\"uhlbacher, and A. Komnik, arXiv:0808.0442.


\bibitem{Werner08} P. Werner, T. Oka and A. J. Millis, ArXiv:0810.2345


\bibitem{Ferretti}
A. Ferretti, A. Calzolari, R. Di Felice, F. Manghi, M. J. Caldas,
M. Buongiorno Nardelli, and E. Molinari,
{\em Phys. Rev. Lett.} {\bf 94}, 116802 (2005).

\bibitem{Darancet}
P. Darancet, A. Ferretti, D. Mayou, and V. Olevano,
{\em Phys. Rev. B} {\bf 75}, 075102 (2007).

\bibitem{Thygesen}
K.S. Thygesen, A. Rubio, {\em J. Chem. Phys.} {\bf 126}, 091101
(2007).

\bibitem{Thygesen07b} 
K.S. Thygesen and A. Rubio, Phys. Rev. B ${\bf 77}$, 115333 (2008).

\bibitem{Thygesen08}
K.S. Thygesen, Phys. Rev. Lett. ${\bf 100}$, 166804 (2008).

\bibitem{Mitra05}
A. Mitra, I.Aleiner and A. J. Millis, Phys. Rev. Lett. {\bf
94} 076404/1-4 (2005).

\bibitem{Mitra07} 
A. Mitra and A. J. Millis, Phys. Rev. B {\bf 76}, 085342 (2007).

\bibitem{Segal07}
D. Segal, D. R. Reichman, and A. J. Millis, Phys. Rev. B {\bf 76}, 195316 (2007).

\bibitem{Anderson}
P.W. Anderson, Phys. Rev. ${\bf 124}$, 41 (1961).

\bibitem{Hedin}
L. Hedin and S. Lundqvist, Solid State Phys. ${\bf 23}$, 1 (1969).

\bibitem{KSS}
B. Kjollerstrom, D.J. Scalpino, and J.R. Schrieffer, Phys. Rev. ${\bf 148}$, 665 (1966).

\bibitem{Hybertsen}
M.S. Hybertsen and S.G. Louie, Phys. Rev. B ${\bf 34}$, 5390 (1986).

\bibitem{Aryasetiawan}
F. Aryasetiawan, O. Gunnarsson O., Rep. Prog. Phys. {\bf 61} 237
(1998), and references therein.

\bibitem{Aulbur00}
W. G. Aulbur, L. Jonsson, and J. W. Wilkins, in {\em Solid State Physics},
edited by H. Ehrenreich and F. Spaepen (Academic, New York, 2000), p. 2, 
and references therein.

\bibitem{Stan06}
A. Stan, N. E. Dahlen and R. Van Leeuwen, {\em Europhys. Lett.} {\bf 76}, 298 (2006).

\bibitem{Schilfgaarde06}
M. van Schilfgaarde, T. Kotani and S. Faleev, {\em Phys. Rev. Lett.} {\bf 96}, 226402 (2006).

\bibitem{Wang}
X. Wang, C.D. Spataru, M.S. Hybertsen and A.J. Millis, Phys. Rev. B,
${\bf 77}$, 045119 (2008).

\bibitem{White}
J.A. White,  Phys. Rev. B ${\bf 45}$, 1100 (1992).

\bibitem{Flex}
N.E. Bickers, D.J. Scalapino and S.R. White, Phys. Rev. Lett. ${\bf 62}$, 961 (1989). 

\bibitem{MeirWingreenLee}
Y. Meir, N.S. Wingreen and P.A. Lee, Rev. Lett. ${\bf 70}$, 2601 (1993).

\bibitem{Hershfield}
S. Hershfield, John H. Davies, and John W. Wilkins,  Phys. Rev. Lett. ${\bf 67}$, 3720 (1991). 

\bibitem{Baranger} 
R. Lui, S.-H. Ke, H. Baranger, and W. Yang, J. Am. Chem. Soc. ${\bf
  128}$, 6274 (2006).

\bibitem{Haldane}
F. D. M. Haldane, J. Phys. C: Solid State Phys. {\bf 11}, 5015 (1978)

\bibitem{Haug}
H. Haug and A.-P. Jauho, {\it Quantum Kinetics in Transport and
  Optics of Semiconductors} (Springer, Berlin, 1996).

\bibitem{Datta}
S. Datta, {\it Electronic Transport in Mesoscopic Systems} (Cambridge, 1995).

\bibitem{notation}
The notation we use throughout Section \ref{section_formalism} (except
for eq. \ref{lead_sigma}) can be easily
generalized to the case of a central 
region with multiple impurity sites and non-overlaping orbitals, 
by replacing the spin index $\sigma$ with
a generalized index $n$ denoting both the site index and the spin
degree of freedom. The generalized notation is used in Appendix B.

\bibitem{Budau}
C. Spataru and P. Budau, J. Phys.: Cond. Matt. {\bf 9}, 8333 (1997).

\bibitem{Spataru}
C.D. Spataru, L.X. Benedict, and S.G. Louie, Phys. Rev. B ${\bf 69}$, 205204 (2004).

\bibitem{Pulay}
P. Pulay, Chem. Phys. Lett. ${\bf 73}$, 393 (1980).

\bibitem{Meir}
Y. Meir and N.S. Wingreen, Phys. Rev. Lett. ${\bf 68}$, 2512 (1992).

\bibitem{diVentra}
M. Di Ventra and S.T. Pantelides,Phys. Rev. B ${\bf 61}$, 16207 (2000). 

\bibitem{T_used}
Throughout subsection IV(A) we consider $k_BT=0$, but plots practically 
undistinguishable from the $k_BT=0$ case 
can be obtained at small non-zero temperatures, such as $k_BT=0.01$.

\bibitem{Spataru_tobepublished}
C.D. Spataru et al., to be published.

\bibitem{Kadanoff}
L.P. Kadanoff and G. Baym, {\it Quantum statistical mechanics : 
Green's function methods in equilibrium and nonequilibrium problems}
(Cambridge, 1989).

\bibitem{Anderson2}
P.W. Anderson, Phys. Rev. ${\bf 164}$, 352 (1967).

\bibitem{Abrikosov}
A.A. Abrikosov, Physics (Long Island City, N.Y.), {\bf 2}, 61 (1965).

\bibitem{Jauho}
A.-P. Jauho, N.S. Wingreen, and Y. Meir, Phys. Rev. B ${\bf 50}$, 5528 (1994).

\end{references}
\end{document}